\input epsf 
\documentstyle[12pt,aasms4]{article}
\def\wig#1{\mathrel{\hbox{\hbox to 0pt{%
          \lower.5ex\hbox{$\sim$}\hss}\raise.4ex\hbox{$#1$}}}}

\newcommand{\beq}{\begin{equation}}
\newcommand{\eeq}{\end{equation}}
\newcommand{\eeql}[1]{\label{eq:#1}\end{equation}}

\newcounter{compteur}

\def\bib{\par\noindent\hangindent=3mm\hangafter=1}

\def\mv{M_V}
\def\msol{M_\odot}

\def\Msol{M_\odot}

\def\te{T_{\rm eff}}

\def\simgr{\,\hbox{\hbox{$ > $}\kern -0.8em \lower 1.0ex\hbox{$\sim$}}\,}
\def\simle{\,\hbox{\hbox{$ < $}\kern -0.8em \lower 1.0ex\hbox{$\sim$}}\,}
\def\wig#1{\mathrel{\hbox{\hbox to 0pt{%
          \lower.5ex\hbox{$\sim$}\hss}\raise.4ex\hbox{$#1$}}}}

\begin{document}

\title{Cepheid models based on self-consistent stellar evolution and
pulsation calculations: the right answer?}
\author{Isabelle Baraffe, Yann Alibert, Dominique M\'era, Gilles Chabrier}
\affil{C.R.A.L (UMR 5574 CNRS), 
 Ecole Normale Sup\'erieure, 69364 Lyon
Cedex 07, France\\ ibaraffe, yalibert, dmera, chabrier @ens-lyon.fr}
\and
\author{Jean-Philippe Beaulieu}
\affil{Kapteyn Institute, PO box 800, 
9700 AV Groningen, The Netherlands\\ beaulieu@astro.rug.nl} 

%%%%%%%%%%%%%%%%%%%%%%%%%%%%%%%%%%%%%%%%%%%%%%%%%%%%%%%%%%%%

\begin{abstract}
We have computed stellar evolutionary models for  stars in a mass range characteristic
of Cepheid variables  ($3<m/\Msol<12$) for different metallicities
representative of the Galaxy and the Magellanic Clouds
populations. The stellar evolution calculations are coupled to a
linear non adiabatic stability analysis to get 
self-consistent mass-period-luminosity relations. The period
- luminosity relation as a function of metallicity is analysed
and compared to the recent EROS observations in the Magellanic
Clouds.
The models reproduce the observed width of
the instability strips for the SMC and LMC.
We determine a statistical P-L relationship, taking into
account the evolutionary timescales and a mass distribution given by a Salpeter mass function.
Excellent agreement is found with the SMC PL relationship
 determined by Sasselov et al. (1997). The models reproduce the change of
slope in the P-L relationship near $P\sim 2.5$ days discovered recently by
the EROS collaboration (Bauer 1997; Bauer et al. 1998) and thus explain this feature in term of stellar
evolution. Some discrepancy, however,
remains for the LMC Cepheids. 

 The models are also in good agreement with Beat Cepheids observed  by the
 MACHO and EROS collaborations.
 We show that most 
%a substantial fraction 
of the 1H/2H Beat 
Cepheids have
not yet ignited central helium burning; they are just evolving off the
Main Sequence toward the red giant branch.
% and are responsible for
%the observed 2H/1H period modes. ***
\end{abstract}

%\newpage
%\twocolumn

\section {Introduction}

Over the past five years, tremendous efforts have been devoted to the
search for dark matter through microlensing effects. This systematic
search has provided two large databases of variable stars in the
Magellanic Clouds due respectively to the EROS group (Renault et al.
1996, Beaulieu \& Sasselov 1996 and references therein) and the 
 MACHO group (Welch et al. 1996 and
references therein). 
%One of the most interesting types of  variable stars 
%is the classical  Cepheids, since they are the most important primary
%distance indicators.
% and form the conerstone of extragalactic distance scales.
The wealth of data collected for Cepheids by both projects provides a unique
tool to analyse  period - luminosity relations and their
sensitivity to metal abundances. A significant number of Beat Cepheids,
oscillating in both the fundamental (F) and first overtone (1H) modes or
the first and second (2H) overtone modes
have also been observed in the Clouds. 
%Beat Cepheids provide
% a crucial test for stellar
%and pulsation theory, since the comparison with observed period ratios
%is free of  uncertainties in the distance modulus or reddening
%and bolometric corrections.

%Although Cepheids have been studied extensively since the discovery by
%Leavitt in 1908 of their fundamental period -
%luminosity (PL) relationship, the
%metallicity dependence of this relation 
%still remains unsettled (see Tanvir 1997 for a
%review). 
The analysis of the EROS Cepheids (Sasselov et al. 1997, S97)
in the small (SMC) and Large (LMC) Magellanic Clouds has recently revived the old debate regarding the metallicity dependence of the period -
luminosity (PL) relationship (see Tanvir 1997 for a
review). A careful comparison of the observed PL relations
in both clouds enabled S97 to disentangle  from
differential
distance and reddening a significant metallicity effect, which is not 
predicted by theoretical analysis (Chiosi, Wood \& Capitano 1993, CWC93). 
%This  bears
%important
%consequences on the determination of the Hubble constant H$_0$ (Beaulieu et
%al.
%1997a). 
%On the theoretical side, the most systematic study 
%has been 
%performed by Chiosi, Wood \& Capitano (1993, CWC93), based on linear
%stability analysis of a large grid of models with different
%composition, effective temperature, mass and mass - luminosity (ML)
%relationships. The metallicity effect on the mass
%luminosity
%relationships was found to be much smaller than the one inferred
% from the EROS observations. However, the models of CWC93 are based on the old %Los Alamos opacity Library (Huebner et al.
%1977), which could be the source of this disagreement. Since then, opacity %calculations have been greatly
%improved, thanks to both the
%Livermore group (OPAL,
%Iglesias \& Rogers 1991; 
%Iglesias, Rogers, \& Wilson 1992) and the OP project (Seaton et al. 1994).

Analysis of Beat Cepheids based on the OPAL opacity data set 
(Iglesias \& Rogers 1991; Iglesias and Rogers 1996) yield now
good agreement for galactic Bump and Beat Cepheids (Moskalik et al. 1992; Christensen-Dalsgaard
1993, Christensen-Dalsgaard and Petersen 1995), but 
discrepancies still remain regarding the Magellanic
Clouds (Christensen-Dalsgaard
\& Petersen 1995; Buchler et al. 1996; Morgan \& Welch 1997, MW97).

%Although
%models using the new opacities yield much better
%agreement with  observed 
%galactic Bump and Beat Cepheids (Moskalik et al. 1992; Christensen-Dalsgaard
%1993, Christensen-Dalsgaard and Petersen 1995), discrepancies still remain %regarding the Magellanic
%Clouds (Christensen-Dalsgaard
%\& Petersen 1995; Buchler et al. 1996).
% An analysis of LMC Beat Cepheids led Christensen-Dalsgaard
%\& Petersen (1995) to conclude that current ML relationships are
%broadly
%consistent with observations if a metallicity spread Z=0.005 - 0.01
%is adopted for the LMC, but no better analysis could be achieved.
%Buchler et al. (1996) found difficulties to reconcile linear stability
%analysis results with the 2:1 resonance at P$_0 \sim 10 $ days
%observed in the LMC and SMC. Moreover, the ML relationships
%inferred from their  analysis
%implies  an unlikely large amount of
%overshooting in stellar evolution calculations. 
%
%More recently, Iglesias and Rogers (1996) have updated the OPAL
%opacities, with an increase of the opacity up to
%20\% for solar-type mixtures. 
%Adopting these improved opacities,
% Morgan \& Welch (1997, MW97) analysed LMC Beat Cepheids, and found that only
%a  ML relationship based on a study of long period galactic Cepheids analysed
%with the Baade-Wesselink method (Simon 1990) and extrapolated in the 
%lower mass regime
%can reproduce
%F/1H and 1H/2H Beat Cepheids. 
%This ML relationship differ significantly
% from stellar
%evolution predictions, indicating a real shortcoming in our
% current understanding of low metallicity evolutionary models or in the
%Beat Cepheid analysis.

None of the afore-mentioned analysis is really consistent since the
stability analysis are based on envelope models using input
physics different from the ones entering  the
stellar evolution calculations. Moreover, the width of the instability
strip and in particular the position of the blue edge are deduced 
from  stability analysis, varying arbitrarily $\te$ for a given 
mass - luminosity (ML) relationship, independently
of any stellar evolution constraint on $\te$.
The aim of
this letter is to present fully consistent stellar and pulsation
calculations based
on the most recent opacity data set. In \S 2 we present the
stellar
and pulsation calculations. Section 3 is devoted to the comparison
with the observed PL relationships for fundamental mode pulsators
in the SMC and LMC. The Beat Cepheids are analysed in
section 4 and conclusions are given in \S 5.

\section {Stellar evolution and pulsation calculations}

Our stellar evolution calculations include  the 
updated OPAL opacities (Iglesias \& Rogers 1996)
for T $> 6000K$.
%based on the Grevesse and Noels (1993) solar
%composition. 
For lower temperatures, we use the
Alexander \& Fergusson (1994) opacities. Convection is
based on  the mixing-length theory 
%(Kippenhahn \& Weigert, 1990), 
with a mixing length parameter
$l_{mix}$ = 1.5 $H_p$. The onset
of convective instability is based on the Schwarzschild criterion, without either
overshooting or
semi-convection prescriptions. 
%Although some of the previous Cepheid analysis
%suggest some amount of overshooting (e.g Buchler et al. 1996),
%we recall that they were based on non-consistent evolution and pulsation
%calculations.
%Therefore, their conclusion may be altered by such inconsistency.
%We indeed prefer, for a first attempt of a comparison of
% self-consistent calculations with observations, to avoid any extra source of %uncertainty
%in the models besides the one
% inherent to the standard mixing length theory.

The linear non-adiabatic stability analysis is based on a radial
pulsation
code originally developed by Lee (1985; Lee and Saio 1989).
% see also Jeannin et al. 1996). 
The convective flux $F_{conv}$ is assumed to be frozen in,
 i.e.  the perturbations $\delta F_{conv}$ are neglected into the linearized
energy equation. 
%even though convection is included in the stellar structures. 
%Note
%that the neglect of convection can alter the structure and
%therefore the stability properties of a model, even in the proximity of
%the blue edge of the instability strip. A test shows that a 4.25
%$\msol$ with Z=0.01, $\te$ = 6200 K and log $\lsol$ = 2.815 has
%a stable fundamental mode with a fully radiative structure,
%but unstable when convection is taken into account. 

%\begin{figure}
%\picplace{2.5cm}
%\epsfxsize=88mm
%\epsfysize=88mm
%\epsfbox{fig1.ps} % where you want to insert a vbox for a figure
%\caption[ ]{ Theoretical H-R diagram for stars with Z=0.004, Y=0.25
%and different masses. Open circles and  crosses indicate the location
%of fundamental unstable modes. The distinction between both symbols
%is explained in the text.
%}
%\end{figure}

The stellar evolution and pulsation codes are coupled. 
%A stability
%analysis
%is then performed consistently on the entire stellar structure
%along evolution,
%from the end of the Main Sequence to the end of the central helium
%burning phase. 
%To save computer time, the pulsation calculation is switched on
% when  $\te$ has varied by 50 K or log L
%by 0.01 along the evolutionary tracks. 
This yields a fully consistent
mass - luminosity - period - evolutionary time relationship.
We have computed stellar models in the mass range 3 - 12 $\msol$ with
various chemical composition (Z, Y) = (0.02, 0.28), (0.01, 0.25)
and  (0.004, 0.25), representative of respectively the galactic,
LMC and SMC environments. 
%Special attention is devoted to the LMC for which
%a variation of Z and Y is taken into account, 
%namely (Z, Y) = (0.01, 0.28) and (0.008, 0.25). 
The detailed evolutionary
tracks will be published elsewhere and we focus in the present
letter on the main results. Figure 1 shows the
evolutionary tracks of Z=0.004 stars in a theoretical
Hertzsprung-Russell diagram (HRD), for m = 3 - 7 $\msol$. Unstable models
are indicated by either open circles or crosses. The distinction
is purely artificial: the crosses refer to the usual Cepheid "blue loop"
instability  which occurs during the central helium burning phase
while the open circles
correspond to the first rapid crossing of the HRD when the stars
evolve off
the Main Sequence toward the red giant branch on a thermal timescale. 
This phase, of typically $\sim 10^4$ years,
is referred to in the following as the ``first
crossing instability''.
%, in opposite to the afore-mentioned usual blue loop instability. 
Since this phase is about
 1/100 shorter than the blue loop one, the large data base now available may
allow the detection of some Cepheids in the first crossing instability.
%We thus take this phase into account in our analysis.
%The position of the red edge, which is theoretically difficult to obtain,
%is determined according to the criterion of CWC93.

As shown in Fig. 1, the extension of the 3 $\msol$
blue loop (crosses, solid line)  is significantly reduced, compared to higher mass stars. 
In this case, the blue edge is determined by the 
{\it turn-over of the evolutionary track}.
This reduction of the blue loop extension is
expected from stellar evolution since blue loops during core
helium burning start at a minimum mass $m_{min}$ and vanish for
m $\simgr 10 \msol$, depending on the initial
composition and the convective mixing treatment (cf.  El Eid 1994). Note that
$m_{min}$ decreases with metallicity. We find $m_{min} \sim 3 \msol$ for
Z=0.004, $m_{min} \sim 3.75 \msol$ for Z=0.01 and  
$m_{min} \sim 4.5 \msol$ for Z=0.02. 
%The very existence and the
%extension of the blue loop are extremely sensitive to the input physics like
%the opacities, the
%convection treatment, etc...This point has been already widely
%investigated
%(see El Eid 1994 and references therein). 
%We note that, according to
%the
%analysis of El Eid (1994) (see also Bressan et al. 1993), 
%core overshooting tends to reduce the extension of the loops. 
   
\section{Period - luminosity relationships}

In Fig. 2a-b we compare  our calculations
with observed fundamental mode pulsators in the LMC and SMC in a P - $M_V$ diagram.
The EROS 
($B_E$, $R_E$) magnitudes were transformed into the 
 Johnson-V magnitude according to Beaulieu et al. (1995). 
We adopt the
extinction E(B-V) = 0.10 for the LMC and  E(B-V) = 0.125 for the
SMC with R$_V$ = 3.3 and
distance moduli $(m-M)_0$ = 18.5 for the LMC and $(m-M)_0$ = 19.13
for the SMC (cf.  Beaulieu et al. 1997a).
Comparison is also
made with the Laney \& Stobie (1994, LS94) data, which 
extend to longer
periods than the EROS data. 
%In fig. 2, we adopt the
%extinction E(B-V) = 0.10 for the LMC and  E(B-V) = 0.125 for the
%SMC with R$_V$ = 3.3 and
% distance moduli $(m-M)_0$ = 18.5 for the LMC and $(m-M)_0$ = 19.13
%for the SMC, as inferred from Beaulieu et al. (1997a). 
%Note that the SMC Cepheids observed by EROS are in the main
%bar and the far arm of the SMC and are not likely to belong to the centroid
%of the SMC, resulting in a distance modulus difference with the 
%LMC by about 0.15 mag larger than with other studies (Caldwell \& Coulson 1986;
%LS94). However, we adopt the same distance modulus for 
%both EROS and LS94 data for the sake of simplicity.
 The transformation of the theoretical quantities (L, $\te$)
into M$_V$ are based on the Allard and Hauschildt (1998) most recent atmosphere
models.
These models originally developed for M-dwarfs do not
extend below log g $<$ 3.5. We therefore use the
bolometric corrections at constant log g = 3.5 but take metallicity
effects into account. 
%Work is under progress
%to calculate atmosphere models and synthetic spectra with lower
%gravities. 
We verified however that gravity effects in the V-band are
 small.  
%The bolometric corrections
%as a function of metallicity given in CWC93,
%although based on different atmosphere models and taking gravity effects into %account, differ at
%most by 0.1 mag compared to our values, 
%confirming the low sensitivity of the (L, $\te$) - M$_V$ calibrations
%on gravity. 

We determine a theoretical statistical P - $\mv$ relationship 
under the form $\mv = A \, {\rm log}P + B$ with a weighted
 least square fit 
by minimizing the quantity :
$$ Q = \sum_i \alpha _i \big( M_{Vi} - A \, {\rm log}P_i - B \big) ^2 $$
where the summation extends over all the stellar masses.
The coefficients $\alpha_i(m_i, t)$ depend on the mass distribution
and on the evolutionary times. We adopt a Salpeter mass function (MF) $dN/dm \propto m^{-2.35}$ 
and the time dependence derives directly
from  the coupled evolution and
pulsation calculations. Although the present day MF in the MC may differ from an initial Salpeter MF, we verified that variations of the slope of the MF between -2 and -4 barely
affect the
slope of the PL relationship. Such a steep MF
favors the lowest mass stars because (i) of the number of stars itself and (ii)
of the
longer time spent in the instability strip as the mass decreases.
 The slope of the PL relationship is thus
hardly affected by stars with m $\ge 7 \msol$.
 %For
%illustration, the time spend by a 7 $\msol$ star in the instability
%strip is at least 100 times smaller than for a 4 $\msol$, for all
%studied metallicities. 
  
In Fig. 2a, comparison is made between the SMC observations and the Z=0.004
models with masses m= 3 - 12 $\msol$. 
%Unstable models in the central
%core He burning phase are indicated by dots. 
The first crossing unstable models
are indicated by open circles but only for m $\le 4 \msol$. 
Although included in the calculations,
the first crossing instability phase for m $\ge 5 \msol$ is statistically
insignificant, since the time spent during this phase is more than
300 times smaller than the time spent by a 4 $\msol$ in the
instability
strip (first crossing and blue loop). We thus predict the existence of
Cepheids with periods $P \simle \,1$ day, as the signature of the first
instability strip, but the observation of these objects requires larger
statistics.
The models agree reasonably well with the observed width of the instability
strip, although they do not reach the observed blue edge of the EROS
data.
%Adopting a lower distance modulus  $(m-M)_0$ = 18.94 and E(B-V) = 0.10
%as given by
%Laney
%and Stobie (1994) would yield a much better agreement,
% except for the faintest objects  ($M_V \ge -1.5, {\rm log} P \le 0.4$). 
The faintest objects  ($M_V \ge -1.5, {\rm log} P \le 0.4$)
seem to
indicate a change of slope in the PL relationship, becoming steeper
for log P $\simle 0.4$, as discovered in the EROS-2 Cepheid sample
and analysed carefully by Bauer (1997) and Bauer et al. (1998).
%Interestingly enough, 
We
note that such a trend is observed in the theoretical relation
near $\sim$3 $\msol$.
%, which has a much reduced blue loop as its higher
This change of slope is thus real and stems from stellar evolution,
illustrating the reduction of the He blue loop as mass
decreases (cf. \S 2).
%*** It reflects both the limited extension and the faintness
%of the blue edge in the HR diagram, which translates into a severely reduced 
%Helium blue loop for the lowest masses, as mentioned above and illustrated
%in Fig. 1. CA ME PARAIT PLUS CLAIR NON ? IL Y A A LA FOIS LE FAIT QUE
%LE BORD BLEU SE RETRACTE, D'OU UN EFFET SUR LA -PERIODE-
%ET LE FAIT QUE LE BORD BLEU DEVIENT TRES FAINT, D'OU UN EFFET SUR LA
%-MAGNITUDE-. ON EST D'ACCORD ? ***
Note also that the period of the minimum mass 
undergoing a blue
loop ($\sim 3 \msol$) is consistent with the faintest observed SMC Cepheids.
Finally the  average  slope of our P - $\mv$ relationship
 is $A$ =  -2.92, in excellent agreement with the slope derived by
S97.  

Fig. 2b shows the results for the LMC with Z=0.01 and masses
from 3.75 to 12 $\msol$. As in Fig. 2a, the models reproduce correctly the
observed width of the instability strip and 
the overall agreement with observations is good.
We note
that 
the minimum theoretical ``unstable'' mass undergoing a blue loop
$m_{min} \sim 3.75 \msol$ for Z=0.01 does not correspond to
the faintest objects observed, which correspond to $\sim 4.25 \msol$. 
% However, we are faced to a real problem trying to reproduce
%a similar slope as  found for the SMC and prescribed  by Sasselov et
%al. (1997). 
We predict that fundamental pulsators with ${\rm log} P \simle 0.3$ days
should be in the first crossing instability.
The slope of the PL relationship with a minimum mass of $4.25 \msol$ is $A$ = -2.50, shallower than the one observed
in the LMC by S97.
In order to test the influence of chemical composition, 
we have recomputed the whole grid of models with (Z=0.008, Y=0.25) and
(Z=0.01, Y=0.28), without any substantial modification of the theoretical slope. %We note
%that 
%the minimum theoretical ``unstable'' mass undergoing a blue loop
%$m_{min} \sim 3.75 \msol$ for Z=0.01 does not correspond to
%the faintest objects observed. 
%This is not due to
%possible uncompleteness of the LMC sample, since the faintest objects
%are still above the detection limit of the EROS experiment.
%This may indicate either the
%influence of non-linear effects, which will change the stability
%properties and may increase the minimum unstable mass, or 
% shortcomings in the stellar models.
%However, the good agreement reached for the SMC indicates that the basic input
%physics is correct. 
 For solar metallicity models,
we find also a slope $A$ = -2.55 shallower than the one
observed in the Milky Way.  A detailed analysis of
the possible sources for such discrepancies is under progress.

%\begin{figure}
%\picplace{2.5cm}
%\epsfxsize=88mm
%epsfysize=88mm
%\epsfbox{fig2a.ps} % where you want to insert a vbox for a figure
%\caption[ ]{ Figure 2a: Period luminosity relationship for SMC. 
%Observations are from EROS (plus) and Laney \& Stobie (1994, full
%diamonds). The models are for Z=0.004 and Y=0.25.
%Thick dots correspond to the He burning unstable models. 
%Open circle corresponds to unstable models in the first crossing
%instability (see text). The slope and minimum mass is indicated
%in the upper left corner}
%
%\end{figure}
%
%\begin{figure}
%\picplace{2.5cm}
%\epsfxsize=88mm
%\epsfysize=88mm
%\epsfbox{fig2b.ps} % where you want to insert a vbox for a figure
%\caption[ ]{ Figure 2b: Same as Fig. 2b for LMC. Models
%are for Z=0.01 and Y=0.25 
%}
%\end{figure}

\section{Beat Cepheids}

Figure 3a-b display the Petersen diagrams for SMC (Fig. 3a), LMC and galactic (Fig.
3b)
Beat Cepheids. The data are taken from  EROS (Beaulieu
et al. 1997b) and  MACHO (Alcock et al. 1995; Alcock et al. 1997) collaborations.
Comparison
is also made with galactic F/1H pulsators (Szabados, 1988).
Agreement
with the F/1H mode pulsators is excellent for the three metallicities,
except for the shortest periods of the LMC objects.
 The first crossing unstable models
are also displayed, but predict slightly too high 1H/F period
ratios and extremely short fundamental periods.   
A variation of Y from 0.25 to 0.28 does not
affect
significantly the period ratios and the Z=0.008
models give slightly higher period ratios than observed in the LMC, by 1-2\%. 
The present models however yield a better
agreement with observations than the one obtained by MW97
using the Chiosi (1990) ML relationship with overshooting.

The most striking and new result shown in Fig. 3 concerns the 1H/2H
pulsators, which are reproduced by  models in the
first crossing instability. Indeed, low luminosities are
required  to reproduce 
the observed period ratios at
very short periods,   as
noted by MW97. The luminosities 
during  the He burning blue loop are already too large and yield
too high P$_1$, as indicated by
the crosses in Fig. 3, as already noted by MW97.
%with ML relationships based on stellar evolution calculations.
In order to
reproduce the observed shortest periods, they need masses m $<$ 2
$\msol$. 
Fig. 1 shows however that the evolution during
the
first crossing takes place at a significantly fainter luminosity and
allows
to reach the shortest periods observed with masses 
m $\ge 3 \msol$. 
%A comparison of the models
% with the SMC Beat Cepheids of Beaulieu et al. (1997b)
% in a P -
%$\mv$
%diagram (fig. 4)
%confirms this result. The first overtone pulsators
%of the EROS sample of single mode Cepheids are also displayed in Fig. 4. 
We find that at least the four faintest 1H/2H
Beat Cepheids of the Beaulieu et al. (1997b) sample can be explained by models
in the first crossing instability. This is statistically reasonable,
considering that 
 the total number
of single mode Cepheids (F or 1H) found by EROS in the SMC is 400. Since 
the lowest masses contribute dominantly to this total number,
the probability to have 4 Beat Cepheids in this sample in
the first crossing instability i.e $4/400 \sim 1\%$ is consistent with
 the ratio of the
time spent in the first crossing instability phase
 to the total time spent in the instability strip (F and 1H
instability
strips included) (cf. \S 2). 

\section {Conclusions}

We have computed self-consistent calculations between
stellar evolution and linear
stability analysis
for SMC, LMC and galactic chemical compositions. 
%The models include the most
%recent OPAL opacities and standard input physics. 
%(the Schwarzschild criterion for the onset of
%convection and no overshooting). 
This
first consistent analysis yields good general agreement with the
observed  P - $\mv$ diagrams  in the SMC and LMC and
reproduces reasonably well the width of the instability strip in both cases.
 
The theoretical slope of the PL relationship is in excellent
agreement with the empirical one derived by S97 for the SMC.
%The minimum mass (3 $\msol$) found to undergo a blue loop during core helium
%burning and thus spending a significant amount of time in the instability
%strip is consistent with the faintest observed single-mode SMC Cepheids. 
The puzzling trend of a change of slope observed for log P $\simle$ 0.4 
and discovered by Bauer (1997) is
reproduced by the models and thus can 
be explained by {\it purely evolutionary} effects due to  the reduced
extension of the blue loop with decreasing mass.
 
For the LMC, the theoretical slope is $\sim$ 15\%
less steep than the observed one. The minimum mass found to undergo
a blue loop ($M = 3.75 \msol, P_{blue\, loop} \sim 2$ days) corresponds to periods shorter 
by 20\% than the ones of the faintest
observed LMC Cepheids ($P_{obs} \sim 2.5$ days). Variations of the metallicity and initial
helium abundance do not solve this discrepancy. This shows the
limitation of the present calculations and points out a remaining
puzzle for the LMC Cepheids. 
%If the minimum periods observed by EROS
%are confirmed by the MACHO observations,
%the reasons of such discrepancy may be
%due either to specifics properties of the LMC (reddening,
%thickness,...), to non-linear effects affecting the pulsation calculations
%or to remaining uncertainties in the current stellar evolution models.
%(convective mixing treatment, 
%mixing length parameter, etc..).
Investigation of 
%these latter possibilities 
remaining uncertainties in the current stellar evolution models
%(convective mixing treatment, 
%mixing length parameter, etc..)
is under progress but requires
numerous lengthy calculations. 
We indeed note that overshooting can change the minimum mass
undergoing a blue loop and thus the minimum period.
However, the good agreement
found for the SMC seems to indicate that the present standard evolutionary
 models are already basically correct.  

%The theoretical predictions for F/1H and 1H/2H Beat Cepheids are in very 
% good agreement with the SMC and LMC observations.
Regarding  Beat Cepheids, 
we show that 1H/2H pulsators with the shortest periods can
{\it only}
be reproduced by stars in the {\it first crossing instability phase}, with
luminosity significantly lower than the one in the usual
blue loop phase. This new result is supported by statistical
arguments (cf. \S 4).
%, since
% the ratio of the time spent in the first crossing
%instability to the total time spent in the instability strip
%is comparable to the ratio of 1H/2H Beat Cepheids to the total number of
%single-mode Cepheids.
 
%For F/1H pulsators, the theoretical period ratios during the first
%crossing
%instability phase are slightly larger than the observed ones.
%These latter 
%For F/1H pulsators, the theoretical period ratios
%are better explained by 
%stellar models undergoing a {\it blue loop} during the core helium burning %phase.
%During this phase, 
%the predicted period ratios are in excellent agreement with the SMC
%Beat Cepheids, but for the LMC, the shortest observed fundamental 
%periods are difficult to reproduce, confirming the puzzle
%of the LMC Cepheids. 
%The fact that  models in the first crossing instability
%predict slightly too high F/1H ratios  but reproduce correctly
%the observed 1H/2H ratios 
%suggests that the
%envelope structure of stars in the first crossing instability is favorable
%to 1H/2H modes but inhibits F/1H oscillations. 
%Only non-linear calculations will be able to confirm these hypothesis
%in the future, when they will
%be able to describe adequately the beat phenomenon.  
% PAS FORCEMENT

Finally, the present letter shows
%shows that standard input physics for evolutionary 
%calculations provide a good description of the basic
%properties of Cepheids. We show 
that consistent calculations between
stellar
evolution and pulsation analysis may give the right answer to some
unexplained features of Cepheids.
 
We are grateful to the EROS group and in
particular to F. Bauer for discussing the 
EROS-2 Cepheid sample prior to publication. We are also grateful
to U. Lee for providing the original
pulsation code and
to W. Glatzel for helpful discussions. The calculations have been performed
on the cray T3E of the Centre d'Etudes Nucl\'eaires de Grenoble. 

\vfill\eject
\section*{References}

\bib Alcock, C., et al. 1995, AJ, 109, 1653
\bib Alcock, C., et al. 1997, preprint astro-ph/9709025
\bib Alexander D. R., and Fergusson, J. W., 1994, \apj, 437, 879
\bib Allard, F., \& Hauschildt, P.H., 1998, \apj, submitted 
\bib Bauer, F. 1997, PhD thesis, Paris VII   
\bib Bauer, F., et al. 1998, \aap, in preparation
\bib  Beaulieu, J.P., et al. 1995, \aap, 301, 137
\bib Beaulieu, J.P., et al. 1997a, \aap, 318, L47
\bib Beaulieu, J.P., et al. 1997b, \aap, 321, L5
\bib Beaulieu, J.P., \& Sasselov, D. 1996,  12th IAP Astr. meeting, "Astrophysical returns of microlensing surveys", eds R. Ferlet and JP. Maillard
%\bib Bressan, F., Fagotto, G., Bertelli, G., \& Chiosi, C. 1993, \apjs, 100, 674
\bib Buchler, J.R., Kollath, Z., Beaulieu, J.P., \& Goupil, M.J. 1996, \apjl,
462, L83
%\bib Caldwell, J.A.R., Coulson, I.M. 1986, \mnras, 218, 223
\bib Chiosi, C. 1990, "Confrontation Between Stellar Pulsation and Evolution",
ASP Conf. Ser., eds. C. Cacciari and G. Clementini, p. 158
\bib Chiosi, C., Wood, P.R., \& Capitano, N. 1993, \apjs, 86, 541 (CWC93)
\bib Christensen-Dalsgaard, J. 1993, "Inside the stars", IAU coll. 137, eds. W.W.
Weiss and A. Baglin, ASP. Conf. Ser., 40, 483
\bib Christensen-Dalsgaard, J., \& Petersen, J.O. 1995, \aap, 299, L17
\bib El Eid, M.F. 1994, \mnras, 275, 983
%\bib Grevesse, N. \& Noels, A., 1993, "Origin and Evolution
%of the Elements", ed. N. Prantzos, E. Vangioni-Flam and M. Cass\'e, p15
%\bib Huebner, W.F., Merts, H.L., Magee, H.N., Jr., \& Argo, M.F. 1977, 
%Los Alamos Sci. Lab. Report LA 6760 MA
\bib Iglesias, C.A., \& Rogers, F.J. 1991, \apj, 371, L73
\bib Iglesias, C.A., Rogers, F.J. 1996, \apj, 464, 943
%\bib Iglesias, C.A., Rogers, F.J., \& Wilson, B. 1992, \apj, 397, 717
%\bib Jeannin, L., Fokin, A.B., Gillet, D., \& Baraffe, I. 1996, \aap, 314, L1
%\bib Kippenhahn, R., \& Weigert, A. 1990, "Stellar Structure and Evolution", %ed. M. Harvit, R. Kippenhahn, V. Trimble and J.P Zahn
\bib Laney, C.D., \& Stobie, R.S. 1994, \mnras, 266, 441 (LS94)
\bib Lee, U. 1985, PASJ 37, 261L
\bib Lee, U., \& Saio, H. 1989, \mnras, 237, 875L
\bib Morgan, S.M., \& Welch, D.L.  1997, AJ, 114, 1183 (MW97)
\bib Moskalik, P., Buchler, R., \& Marom, A. 1992, \apj, 385, 685
\bib Renault, C., et al. 1996, 12th IAP Astr. meeting, 
%"Astrophysical returns of
%microlensing surveys", 
eds R. Ferlet and JP. Maillard
\bib Sasselov, D., et al. 1997, \aap, 324, 471 (S97)
%\bib Seaton, M.J., Yan., Y., Mihalas, D., \& Pradhan, A.K. 1994, \mnras, 266, %805
%\bib Simon, N.R. 1990, "Confrontation Between Stellar Pulsation and Evolution",
%ASP conf. Ser., eds C. Cacciari and G. Clementini, p. 193
\bib Szabados, L. 1988, "Multimode Stellar Pulsations", eds. G. Kovacs, L. Szabados and B. Szeidl, p. 1
\bib Tanvir, N.R. 1996, Proceedings of the STScI May Symposium, "the 
Extragalactic Distance Scale, eds. M. Livio, M.Donahue and N. Panagia, p. 91
\bib Welch, D.L, et al. 1996,  12th IAP Astr. meeting, 
%"Astrophysical returns of
%microlensing surveys", 
eds R. Ferlet and JP. Maillard

\vfill\eject

%\centerline{\bf Figure Captions}

\begin{figure}
\centerline{\bf Figure Captions}
\par
%\picplace{2.5cm}
%\epsfxsize=150mm
%\epsfysize=150mm
%\epsfbox{fig1.ps} % where you want to insert a vbox for a figure
%\caption[ ]{
Fig. 1.-- Theoretical H-R diagram for stars with Z=0.004, Y=0.25
and different masses. Open circles and  crosses indicate the location
of fundamental unstable modes. The distinction between both symbols
is explained in the text (\S 2).
\end{figure}

\begin{figure}
%\picplace{2.5cm}
%\epsfxsize=150mm
%\epsfysize=150mm
%\epsfbox{fig2a.ps} % where you want to insert a vbox for a figure
%\caption[2a ]{
Fig. 2a.-- Period- magnitude relationship for the SMC. 
Observations are from EROS (plus) and Laney \& Stobie (1994, crosses). The models are for Z=0.004 and Y=0.25.
Dots correspond to the core He burning unstable models. 
Open circles correspond to unstable models in the first crossing
instability phase (see text). The straight line is the average statistical
P - $\mv$ relation derived from the present models (see text).
The slope and minimum mass are indicated
in the upper left corner.
\end{figure}

\begin{figure}
%\picplace{2.5cm}
%\epsfxsize=150mm
%\epsfysize=150mm
%\epsfbox{fig2b.ps} % where you want to insert a vbox for a figure
%\caption[2b ]{
Fig. 2b.-- Same as Fig. 2a for LMC data and models
with Z=0.01 and Y=0.25 
%}
\end{figure}

\begin{figure}
%\picplace{2.5cm}
%\epsfxsize=150mm
%\epsfysize=150mm
%\epsfbox{fig3a.ps} % where you want to insert a vbox for a figure
%\caption[ ]{
Fig. 3a.-- Petersen diagram for SMC models with
Z=0.004. Observations are from MACHO (full circles) and EROS (full
squares).
Crosses correspond to unstable models during the He burning blue loop. 
Open circles correspond to unstable models in the first crossing
instability (see text). The corresponding masses are indicated for
both F/1H (lower sequence) and 1H/2H (upper sequence)
mode pulsators.
%}

\end{figure}

\begin{figure}
%\picplace{2.5cm}
%\epsfxsize=150mm
%\epsfysize=150mm
%\epsfbox{fig3b.ps} % where you want to insert a vbox for a figure
%\caption[ ]{
 Fig. 3b.-- Same as Fig. 3a for LMC models with Z=0.01
 and Y=0.25. The F/1H galactic pulsators of Szabados (1988) are added
(full diamonds) and compared to Z=0.02 models (the lowest sequence,
+). The corresponding masses are indicated for Z=0.01. The Z=0.02
models are only shown during  the He blue loop and 
 correspond to 4.5 to 6 $\msol$ from left to right.
%}
\end{figure}

%\begin{figure}
%\picplace{2.5cm}
%\epsfxsize=150mm
%\epsfysize=150mm
%\epsfbox{fig4.ps} % where you want to insert a vbox for a figure
%\caption[ ] { 
%Fig. 4.-- $\mv$ as function of the first overtone period
%of single mode Cepheids (plus) and 1H/2H beat Cepheids (full
%squares) from EROS observations. We use a distance modulus (m-M)$_0$ = 18.94
%and reddening E(B-V) = 0.10.
%The symbols are the same as in Fig. 2.
%The first crossing unstable models (open circles) 
%are displayed for 3, 3.25, 3.5 and
%4 $\msol$, from left to right.
%\end{figure}

%\end{document}
\vfill\eject

\begin{figure}
\epsfbox{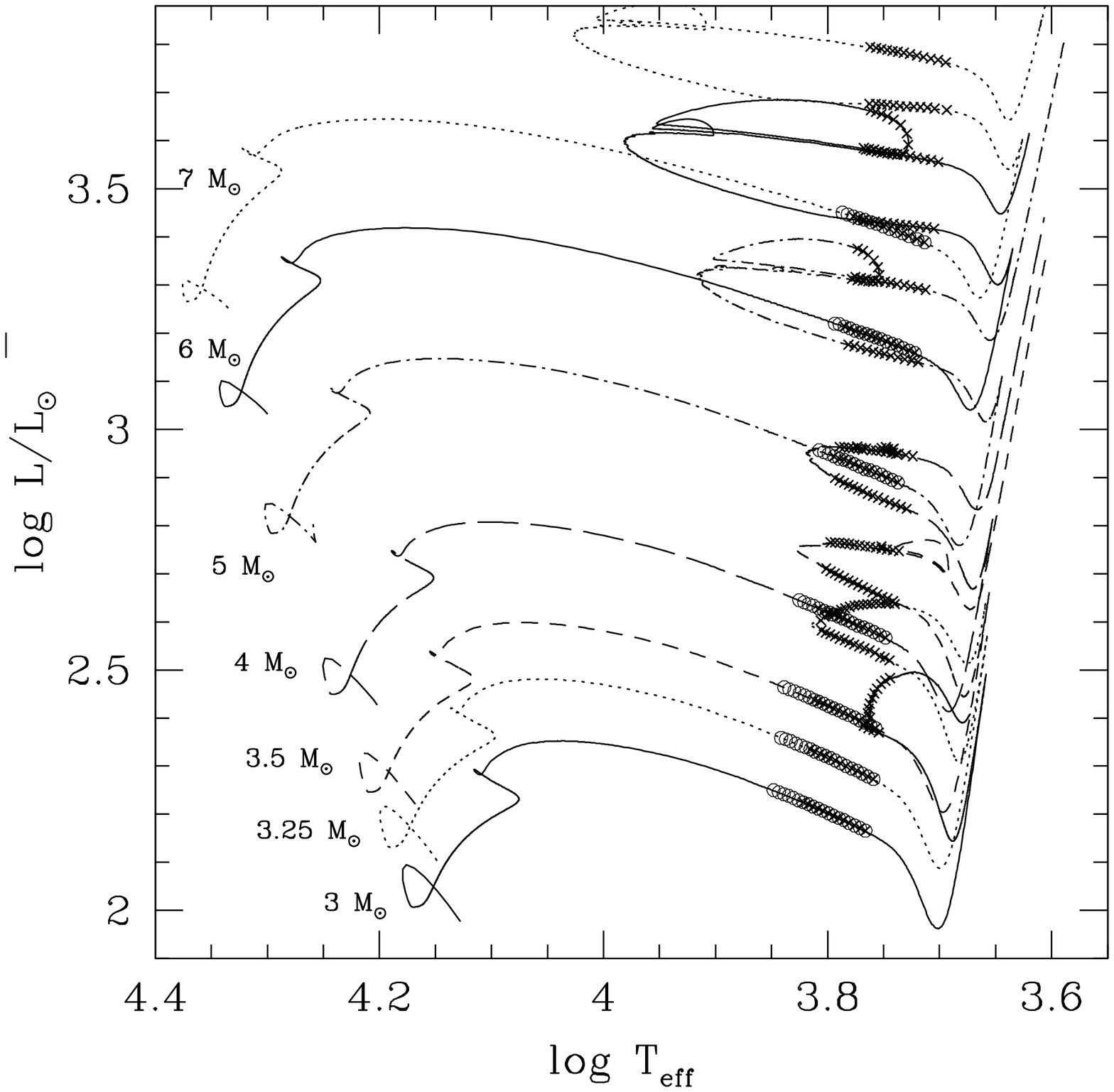}
\end{figure}

\begin{figure}
\epsfbox{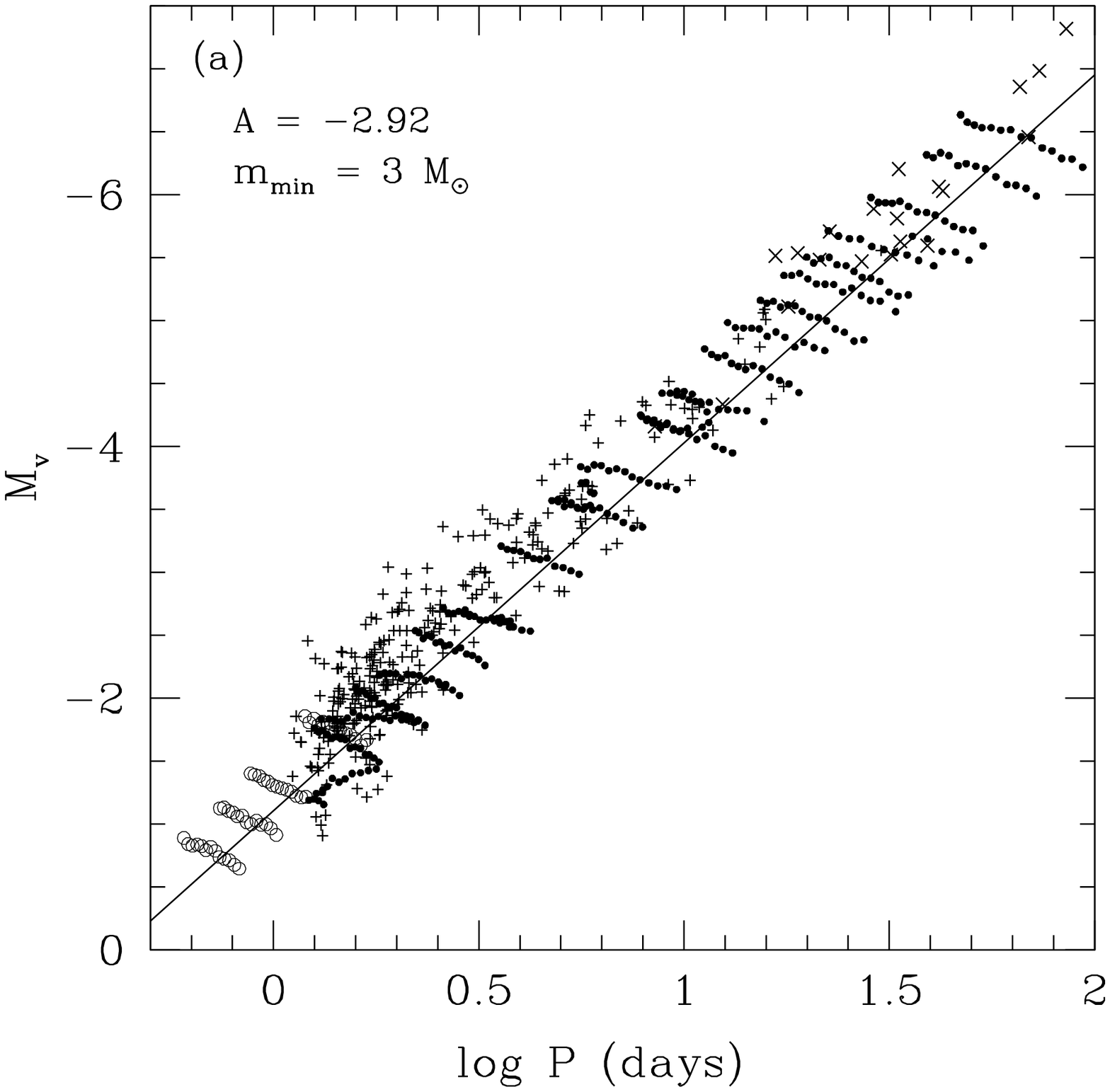}
\end{figure}

\begin{figure}
\epsfbox{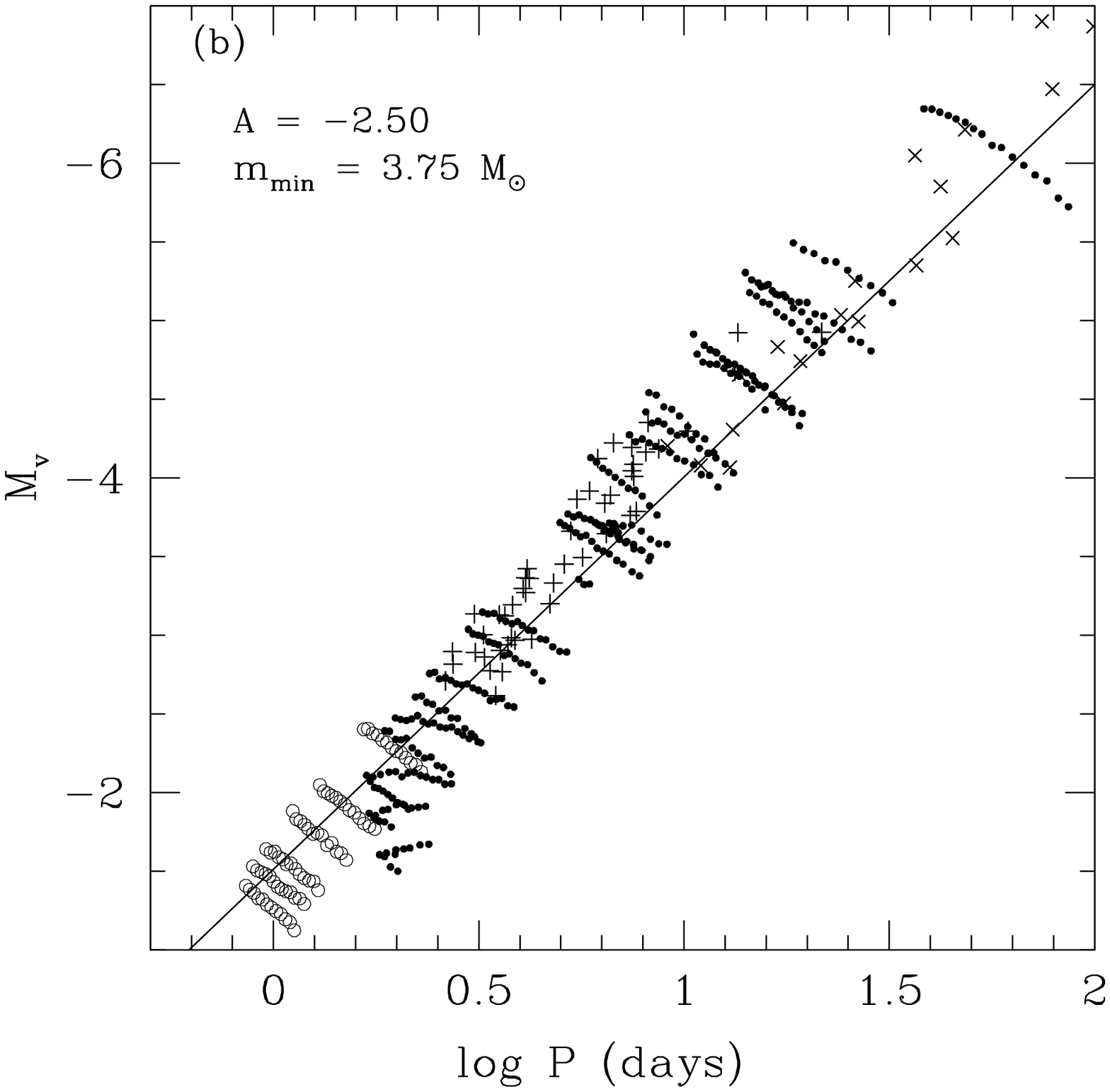}
\end{figure}

\begin{figure}
\epsfbox{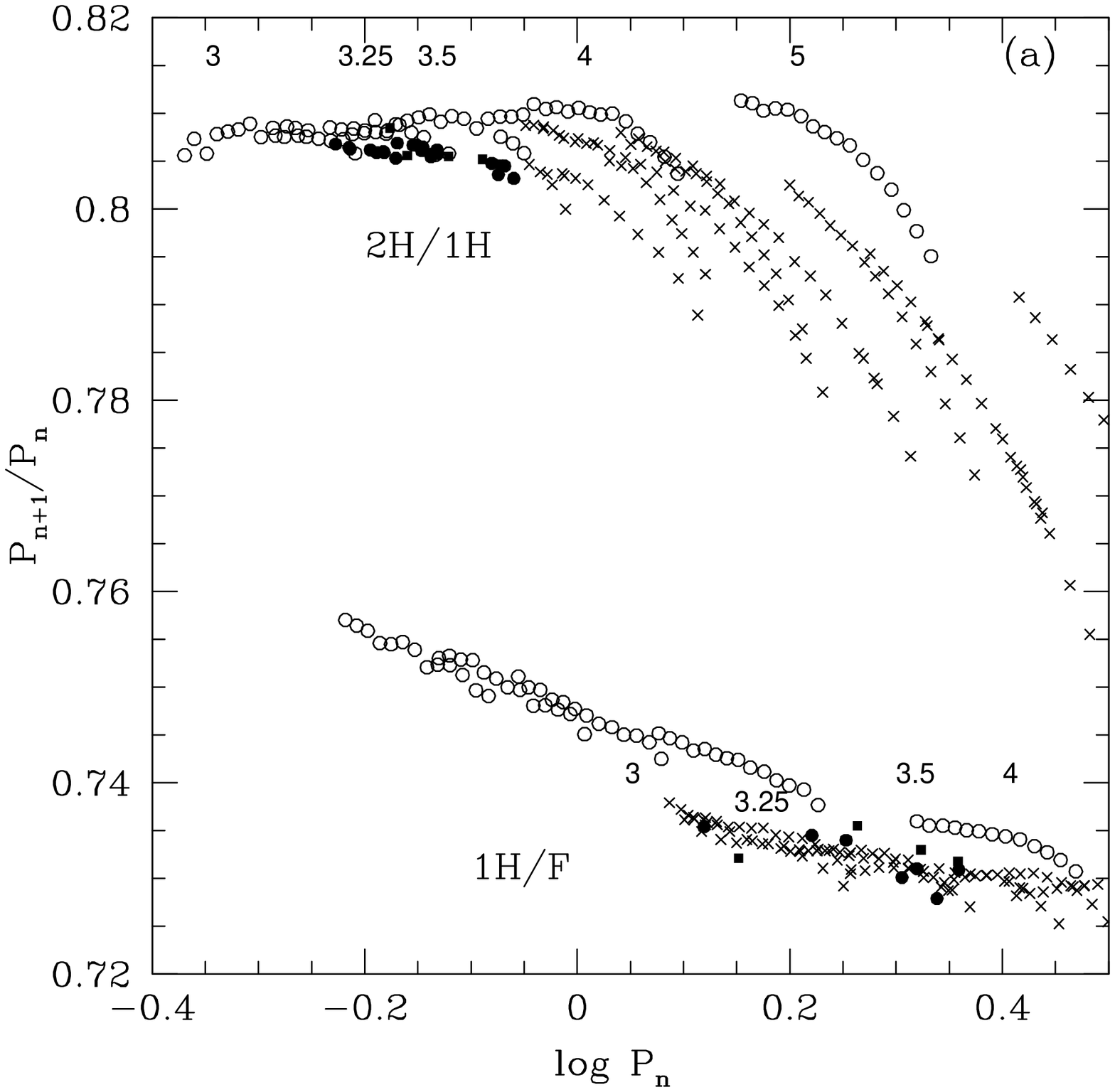}
\end{figure}

\begin{figure}
\epsfbox{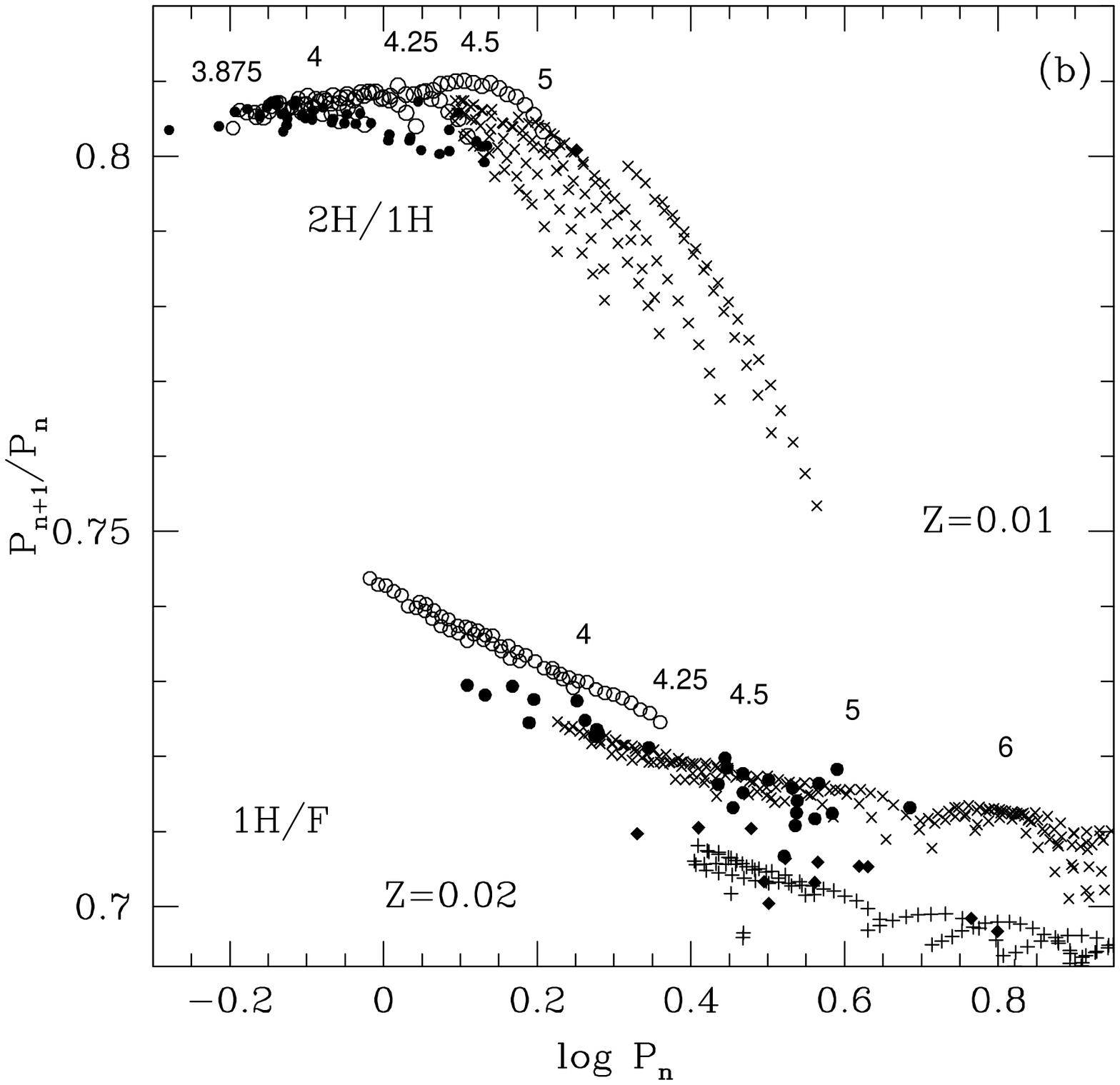}
\end{figure}

\end{document}